\documentstyle[11pt]{article}
\def\fnote#1#2{\begingroup\def\thefootnote{#1}\footnote{#2}\addtocounter
{footnote}{-1}\endgroup}

\begin{document}

\hfill{UTTG-06-08}

\vspace{36pt}

\begin{center}
{\large {\bf {Non-Gaussian Correlations Outside the Horizon}}}

\vspace{36pt}
Steven Weinberg\fnote{*}{Electronic address:
weinberg@physics.utexas.edu}\\
{\em Theory Group, Department of Physics, University of
Texas\\
Austin, TX, 78712}

\vspace{30pt}

\noindent
{\bf Abstract}
\end{center}
\noindent
It is shown that under essentially all conditions, the non-linear classical equations governing gravitation and matter in cosmology have a solution in which far outside the horizon in a suitable gauge the reduced spatial metric (the spatial metric divided by the square of the Robertson--Walker scale factor $a$) is time-independent, though with an arbitrary dependence on co-moving coordinates, and all perturbations to the other metric components and to all matter variables vanish, to leading order in $1/a$.  The corrections are of order $1/a^2$, and are explicitly given for the reduced metric in a multifield model with a general potential.  Further, this is the solution that describes the metric and matter produced by single-field inflation.  These results justify the use of observed non-Gaussian correlations (or their absence) as a test of theories of single-field inflation, despite our ignorance of the constituents of the universe while fluctuations are outside the horizon after inflation, as long as graphs with loops can be neglected.

 \vfill

\pagebreak

\begin{center}
{\bf I. Introduction}
\end{center}

Non-Gaussian cosmological correlations are attracting increasing interest as an observational test of detailed theories of inflation[1].  But there is a problem in calculations of observable non-Gaussian correlations.  Given any specific Lagrangian for the scalar fields that play a role in inflation, we know in principle how to calculate the  correlation functions for these fields and gravitation up to the end of inflation.  And given any set of correlation functions for gravitational and matter and radiation perturbations at some time in the relatively recent era when the temperature is well below the QCD scale,  we know enough about the contents of the universe to calculate the subsequent evolution of these correlations.    But in the intervening era the universe went through a sequence of transformations about which we know almost nothing,  including reheating, lepton and baryon synthesis, and cold matter decoupling.  So how can we
use assumptions about inflation to calculate observable correlations?

The only thing that gives us any hope in understanding cosmological correlations  is that the wavelengths at which these correlations are observed were outside the horizon during the whole period from a time during inflation until a relatively recent time when the contents of the universe are reasonably well understood.
But in order to take advantage of this fact, we need to identify a set of variables whose correlation functions are time-independent for wavelengths outside the horizon.  In the linear approximation, it is known that quantum fluctuations in single-field inflation produce adiabatic fluctuations, in which the curvature perturbation  $\zeta$ as well as the amplitude of gravitational waves become constant outside the horizon[2].  This is enough to show that Gaussian correlations of these quantities remain  
time-independent after inflation, as long as the wavelength is outside the horizon, at least when quantum effects are neglected.  But for non-Gaussian correlations we must work with the full non-linear field equations.

It is known that the classical non-linear field equations for single-field inflation have an ''adiabatic'' solution for which $\zeta$ and the gravitational wave amplitude  become time-independent at late times during inflation[3], and it has further been shown[4] that these quantities become time-independent outside the horizon both during and after single-field inflation, but this is only part of the story.  To provide initial conditions for a calculation of fluctuations through horizon re-entry and  until the present, we need to know  not only $\zeta$ and the other metric components but also  the matter (including radiation) perturbations before horizon re-entry.    It is sometimes taken as part of the definition of the adiabatic solution that (in a suitable gauge) these matter perturbations vanish, but it needs to be shown that such a solution exists, and that in some circumstances the universe is described by this solution.

Section II of this paper shows that whatever the constituents of the universe and the classical equations governing them may be, these equations have  a solution for which in a suitable gauge,  as long as all relevant wavelengths are sufficiently far outside the horizon,   all components of the reduced metric $\tilde{g}_{ij}\equiv g_{ij}/a^2$ become  time-independent functions of position; $g_{00}$ becomes $-1$;  $g_{i0}$ vanishes; and all matter densities, pressures, and velocities become equal to their unperturbed values; in all cases with corrections  of order $(k/aH)^2$.  (As usual, $a(t)$ is the Robertson--Walker scale factor and $H(t)\equiv \dot{a}(t)/a(t)$, while $k$ is the largest relevant wave number.)    The argument for these results is based on  considerations of broken symmetry, similar to those used to derive the form of the chiral Lagrangian for soft pions[5].  It relies only on the general covariance of the underlying equations, and the usual assumption that these equations have a solution of the Robertson--Walker form.  This argument may be regarded as a substitute for a ``separate universe'' assumption[6], but it  gives more detailed information, and some may find it more convincing.  

In Section III we verify these results in  a fairly general model of matter fields, in a special gauge that allows the calculation of metric perturbations outside the horizon without at the same time having to solve the equations for matter perturbations.  In particular we confirm that in this model there is always an adiabatic solution in which the corrections to the leading terms for both the metric and the matter variables are of order $(k/aH)^2$.  This much is already known for the metric components[4], but here we also  obtain explicit results\fnote{**}{{\em Added note}: In a paper now in preparation, I show that these explicit results are not limited to scalar field theories, but are quite general for theories with  anisotropic inertia and vorticity that vanish to order $a^{-2}$.}   for the form of the $O(1/a^2)$ and $O(1/a^3)$ terms in $\tilde{g}_{ij}$[7].  In general this is only one of many possible solutions; whether or not it is the solution that describes the real world depends on the details of inflation.  In Section IV we show that single field inflation in this model leads to matter and metric fields that are described by the adiabatic solution after inflation, as long as all wavelengths are outside the horizon.

There is a problem with all such classical field calculations.  Even if we could show that under all circumstances the full non-linear classical field equations have a solution for which $\zeta$ and gravitational wave amplitudes become constant outside the horizon, and that the initial conditions provided by single-field inflation (or a state of thermal equilibrium after inflation) produce perturbations that are described by this solution, we still would not know that Heisenberg picture quantum operators for  $\zeta$ and gravitational wave amplitudes become constant at late times, and so we would not know that the correlation functions become constant outside the horizon.  ``Outside the horizon'' means that for {\em all} relevant co-moving wave numbers $k$, we have $k/a\ll H$.  For quantum operators there can be no clear meaning to this, because whatever the wave numbers  at which the correlation functions are observed, quantum fluctuations can carry arbitrarily large virtual wave numbers.  That is, there is no limit on the wave numbers circulating in the loops in general graphs, however small we make the wave numbers for the external lines of these graphs.   For inflation with a single inflaton field, the loop contributions to the correlation function of $\zeta$  are  integrals over virtual wave numbers $p$ with  integrands that are time-independent when the virtual as well as the external wavelengths are outside the horizon, but virtual wavelengths can not be constrained to remain outside the horizon, because the integrals over virtual wave numbers are ultraviolet-divergent.  (For examples, see [8].)  True, we can assume that the ultraviolet divergence is canceled by counterterms arising from $\sqrt{-{\rm Det} g}\,R^{\mu\nu}R_{\mu\nu}$ and $\sqrt{-{\rm Det} g}\,R^2$ terms in the Lagrangian density[9], but if this cancelation results in an effective cut-off at $p/a$ of order $H$, then the correlation functions will involve powers of $\ln a$[8].  Detailed calculations[10] confirm the presence of such time-dependent corrections in loop contributions to correlation functions.

Fortunately, in many theories the tree graphs make much larger contributions to the correlation functions than loop graphs.  For a tree graph the wave number associated with any internal line is just a sum of wave numbers of several external lines, so all internal wavelengths can be assumed to be outside the horizon if the external wavelengths are.  Thus at least in some theories, one can treat the non-linear field equations as if all relevant wavelengths were outside the horizon, and hope that quantum effects do not introduce large corrections.  Alternatively, one can limit oneself to the tree approximation from the beginning, and assume that tree graphs give a good approximation to the correlation functions.

A recent paper[11] showed how to calculate the sum of tree graphs for the generating function for general correlation functions by solving the classical equations of motion subject to certain constraints that depend on the current appearing in the generating function.  This is reviewed here in an appendix, using a simplified notation and adding some necessary comments.  In order to conclude in this formalism that correlation functions become time-independent outside the horizon, it is not enough to show that the solution of the non-linear classical field equations becomes time-independent at late times during inflation.  As reviewed here in the appendix, one must also show that the effect of the constraints that are imposed at the time at which the correlation functions are measured becomes independent of this time when all wavelengths are outside the horizon, and also that a certain integral converges.  In Section V we show that these conditions are all satisfied during and after single-field inflation for the quantity $\tilde{g}_{ij}\equiv g_{ij}/a^2$.  (The same argument applies to any function of $\tilde{g}_{ij}$, such as the quantity $\zeta\equiv \ln\sqrt{{\rm Det}\tilde{g}}$ studied in [3].)         Thus in order for parametric amplification during reheating[12] to produce significant changes in the correlation functions, such effects would have to amplify perturbations by a factor of order   $e^{120}$ to $e^{140}$.  

In summary, these results provide a practical program for calculating observable correlation functions from theories of single-field inflation.   

\vspace{6pt}

\noindent
(i) First calculate the correlation functions of $\tilde{g}_{ij}\equiv g_{ij}/a^2$ (or any functions of $\tilde{g}_{ij}$) sufficiently late after horizon exit during inflation so that they are time-independent, using a definition of the time  coordinate for which the inflaton field is unperturbed.  (This would presumably be done by direct calculation of tree graphs, as already done in [3] for the bispectrum, rather than by using the methods of Section V, which are intended only to provide a proof of the time-independence  outside the horizon of the sum of tree graphs for any correlation function of $\tilde{g}_{ij}$.)  If we like we can separate a curvature perturbation $\zeta$ by following [3] and writing
$$ \tilde{g}_{ij}=e^{2\zeta}[e^\gamma]_{ij}\;,~~~~~~~{\rm Tr}\,\gamma=0\;,
$$ or alternatively by writing
$$ \tilde{g}_{ij}=\delta_{ij}+2\zeta\,\delta_{ij}+\gamma_{ij}\;,~~~~~~~{\rm Tr}\,\gamma=0\;.$$
These different definitions of course give different non-Gaussian correlation functions for $\zeta$, but with either definition the correlation functions are constant outside the horizon.

\vspace{6pt}

\noindent
(ii) At a time which is sufficiently early so that the all wavelengths  are still outside the horizon, but late enough so that the contents of the universe are  well understood, take the correlation functions of $\tilde{g}_{ij}$ or of functions of $\tilde{g}_{ij}$ to be given by the results of (i), and take all correlation functions involving $g_{i0}$ and/or $g_{00}+1$ and/or matter or radiation perturbations to vanish.

\vspace{6pt}

\noindent
(iii)  Use the results of (ii) as initial conditions for calculation of the subsequent evolution of the correlation functions for gravitational and matter and radiation perturbations when the wavelengths re-enter the horizon.  This can be done by using the classical field equations to derive coupled differential equations for the various correlation functions, but such calculations are outside the scope of this paper.

\vspace{6pt}

\noindent
The above program appears to be more or less what is done in recent work on non-Gaussian correlations[13].  The aim of this paper is to clarify the justification for these calculations.

\begin{center}
{\bf II. The General Adiabatic Solution}
\end{center}

We assume, as usual in cosmology, that whatever the dynamical equations governing the metric and matter (including radiation) variables may be, these equations have a solution in which the metric takes the Robertson--Walker form, with $g_{00}=-1$, $g_{0i}=0$, and $\tilde{g}_{ij}\equiv g_{ij}/a^2 =\delta_{ij}$, and in which all matter variables take their unperturbed form; that is, all densities and pressures and scalar fields are functions only of time, and all velocities and other 3-vectors vanish.  This section will give a very general argument that, whatever the constituents of the universe and the generally covariant equations governing them and the metric may be, for a suitable choice of spacetime coordinates, these equations  always also have a family of solutions that we will call adiabatic, which for large $a(t)$ have the following properties:
\begin{enumerate}
\item The metric for any of these solutions has components with 
\begin{eqnarray}
&&g_{00}({\bf x},t)=-1+O\Big(a^{-2}(t)\Big)\;,~~~g_{i0}({\bf x},t)=O\Big(a^{-2}(t)\Big)\;,\nonumber\\&&~~~g_{ij}({\bf x},t)=a^2(t)\left[{\cal G}_{ij}({\bf x})+O\Big(a^{-2}(t)\Big)\right]\;,
\end{eqnarray}
where ${\cal G}_{ij}({\bf x})$ is an arbitrary function only of the spatial coordinates.  (Different choices of this function  characterize the different members of this family of solutions.)
\item Whether or not the energy and momentum of any particular constituent of the universe is separately conserved, its energy-momentum tensor has the form
\begin{eqnarray}
&&T_{00}({\bf x},t)=\bar{\rho}(t)+O\Big(a^{-2}(t)\Big)\;,~~~T_{i0}({\bf x},t)=O\Big(a^{-2}(t)\Big)\;,\nonumber\\&&~~~~~T_{ij}({\bf x},t)=a^2(t)\left[{\cal G}_{ij}({\bf x})\bar{p}(t)+O\Big(a^{-2}(t)\Big)\right]\;.
\end{eqnarray}
(Here and below, a bar over any quantity indicates its unperturbed value.)
\item
Any four-scalar  $s({\bf x},t)$, such as a temperatures, number density, or scalar field, has the form
\begin{equation}
s({\bf x},t)=\bar{s}(t)+O\Big(a^{-2}(t)\Big)\;.
\end{equation}
\end{enumerate}
We are not assuming a  de Sitter expansion, but in counting powers of $1/a$, we shall take $H$ and its time derivatives to be of zeroth order in $a$, so that quantities like $\dot{a}(t)$ and $\int a(t)\,dt$ are counted as being of first order in $a$.   It should be understood that since the scale of $a$ has a physical significance only when $a$ multiplies a co-moving coordinate, it follows that when we calculate correlation functions with a typical co-moving wave number $k$, a factor  $a^{-1}$ will always be accompanied with a factor  $k$.  Since $k/a$ has the same dimensions as $H\equiv \dot{a}/a$,  we can anticipate that the dimensionless parameter that characterizes the smallness of a term of order $a^{-n}$ is $(k/aH)^n$. Thus this theorem gives good approximations to the adiabatic solutions both after horizon exit during inflation, when $k/aH$ is decreasing, and before horizon re-entry after inflation, when $k/aH$ is increasing, as long as $k/aH$ is sufficiently small.

Of course, these solutions are in general far from unique, and the statement that these adiabatic solutions exist does  not tell us that one of these solutions actually describes the metric and matter of the universe.  As we will see in  section IV,  if we start with single-field inflation then the universe will thereafter be described by an adiabatic solution.  Also, even when the universe is described by an adiabatic solution, we need a detailed model of inflation to calculate the function ${\cal G}_{ij}({\bf x})$ in Eq.~(1).

To prove the existence of the adiabatic solutions, we will make use of an argument based on  the broken symmetry of general covariance.  As already mentioned, we are assuming that the dynamical equations have a solution in which the metric takes the Robertson--Walker form,  and in which all matter variables take their unperturbed form, with pressures and densities only functions of time, and vanishing co-moving velocities.    Now, whatever they are, the dynamical  equations will be invariant under all coordinate transformations, but this solution is not.  In particular, if we subject the space coordinates to a matrix transformation $x^i\rightarrow x'^i=A^i{}_jx^j$, with $A^i{}_j$ an arbitrary constant real matrix, then we get another exact solution, with $g_{00}=-1$, $g_{i0}=0$,  but now with  $\tilde{g}_{ij}\equiv g_{ij}/a^2$ equal to the arbitrary constant positive real matrix $(A^TA)_{ij}$.  The energy-momentum tensor of any constituent of the universe (whether or not separately conserved) will  in the new coordinate system still have the perfect fluid form, $T_{\mu\nu}=g_{\mu\nu}\bar{p}+\bar{u}_\mu \bar{u}_\nu(\bar{p}+\bar{\rho})$, with the same density $\bar{\rho}(t)$, pressure $\bar{p}(t)$, and velocity $\bar{u}_i=0$, $\bar{u}_0=-1$, but now with the new metric.  

Instead of this exact solution, now consider what we will call a ``trial configuration'' in which $g_{00}=-1$, $g_{i0}=0$ and all densities, pressures, and velocities are unperturbed, but  with  $\tilde{g}_{ij} $ an  arbitrary time-independent positive matrix function ${\cal G}_{ij}({\bf x})$ of the co-moving space coordinates $x^i$, {\em not necessarily close to} $\delta_{ij}$.  This trial configuration  is of course not a solution of the field equations, but since it {\em would} be a solution if $\tilde{g}_{ij}$ were constant, it fails to be a solution only because there are terms in the field equations in which space derivatives act on $\tilde{g}_{ij}$.   (Up to this point, this is just like the argument used to derive the effective chiral Lagrangian for soft pions[5].)  The spatial derivatives of ${\cal G}_{ij}({\bf x})$ thus act as forcing terms, that drive the actual solution away from the trial configuration.  That is, making the tentative assumption that the differences between metric or matter variables and their values in the trial configuration are small perturbations when $a(t)$ is sufficiently large, these perturbations satisfy a set of coupled inhomogeneous linear differential equations, with  left-hand sides that are linear combinations of time derivatives of these perturbations, and  right-hand sides that involve spatial derivatives of ${\cal G}_{ij}({\bf x})$.  We will see concrete examples of such equations in the next section.

Now, as a special case of general covariance, the field equations must be invariant under the substitution $x^i\rightarrow \lambda x^i$ (with $\lambda$ an arbitrary constant) if we also subject other quantities to appropriate transformations: 3-tensors such as $g_{ij}$ and  $T_{ij}$ transform as $g_{ij}\rightarrow \lambda^{-2}g_{ij}$ and $T_{ij}\rightarrow \lambda^{-2}T_{ij}$, while 3-vectors such as  
$g_{i0}$ and  $T_{i0}$ transform as $g_{i0}\rightarrow \lambda^{-1}g_{i0}$ and $T_{i0}\rightarrow \lambda^{-1}T_{i0}$. (Here $T_{\mu\nu}$ may be the energy-momentum tensor of any one constituent of the universe, even if not separately conserved.)  It is convenient to express this as invariance under a scale transformation:
\begin{equation}
x^i\rightarrow \lambda x^i\;,~~~~~~~~~a(t)\rightarrow \lambda^{-1}a(t)\;,
\end{equation}
that leaves invariant various reduced quantities
\begin{eqnarray*}
&\tilde{g}_{ij}({\bf x},t)\equiv  g_{ij}({\bf x},t)/a^2(t)\;,~~~~~~&\tilde{g}_{i0}({\bf x},t)\equiv  g_{i0}({\bf x},t)/a(t)\;,\\
&\tilde{T}_{ij}({\bf x},t)\equiv  T_{ij}({\bf x},t)/a^2(t)\;,~~~~~~&\tilde{T}_{i0}({\bf x},t)\equiv  T_{i0}({\bf x},t)/a(t)\;,
\end{eqnarray*}
as well as all 3-scalars such as temperature, densities, scalar fields, and also  $g_{00}$ and $T_{00}$.
The forcing term for any scale-invariant perturbation must be scale-invariant, and three-dimensional coordinate invariance requires it to have the same transformation under purely spatial coordinate transformations as the perturbation, so the perturbation of any scale-invariant quantity away from its value in the trial configuration will be proportional to as many powers of $1/a$ as appear in the scale-invariant quantity formed from $1/a$ and derivatives of ${\cal G}_{ij}$ that has the same transformation property under three-dimensional coordinate transformations as the perturbation in question.  For perturbations to the scale-invariant quantities $g_{ij}/a^2$ or $T_{ij}/a^2 $, the scale-invariant forcing terms with the minimum number of factors of $1/a$  are proportional to the  3-tensors  ${\cal R}_{ij}/a^2$ or ${\cal G}_{ij}{\cal R}/a^2$, where ${\cal R}_{ij}$ is the 3-dimensional Ricci tensor for the 3-metric ${\cal G}_{ij}$, and ${\cal R}= {\cal G}^{kl}{\cal R}_{kl}$, with ${\cal G}^{ij}$  the reciprocal of ${\cal G}_{ij}$.  Likewise,  for perturbations to  the scale-invariant quantities $g_{i0}/a$ or $T_{i0}/a$, the scale-invariant forcing terms with the minimum number of factors of $1/a$  are proportional to the scale-invariant 
3-vectors $\partial_i{\cal R}/a^3$, and  for perturbations to  the scale-invariant quantities $g_{00}$ or $T_{00}$, the forcing terms with the minimum number of factors of $1/a$  are proportional to the scale-invariant 3-scalar ${\cal R}/a^2$, and likewise for perturbations to any other scale-invariant 3-scalar.  Thus the difference between the values of  the quantities $g_{ij}/a^2$, $T_{ij}/a^2$, $g_{i0}$, $T_{i0}$, $g_{00}$, $T_{00}$, and 3-scalars like temperatures or scalar fields and the values of the corresponding quantities in the trial configuration are all of order $1/a^2$, as was to be proved.  In particular, all these perturbations are small for sufficiently large $a(t)$,  as tentatively assumed in proving the existence of these solutions.

The general solution for the perturbations to the trial configuration consists of a sum of the solution of the inhomogeneous differential equations, with derivatives of ${\cal G}_{ij}({\bf x})$ as forcing terms, plus solutions of the corresponding homogeneous equations.  In a completely general theory of inflation the solutions of the homogeneous equation could have any magnitude.  However, we will see in Section IV that in single field inflation they are also of order $a^{-2}$.

The form (1), (2), for the adiabatic solutions is not valid for all choices of spacetime coordinates, but it is easy to impose gauge-fixing conditions on the   coordinates that are consistent with this form.  We can choose the time-coordinate so that any one three-scalar, such as a scalar field or the temperature, is unperturbed.  (A generalized version of this choice of gauge is adopted in Section III.)  To choose the space coordinates, we note that under a time-dependent transformation $x^i\rightarrow x'^i({\bf x},t)$ that leaves the time invariant, the metric component $g^{i0}$ undergoes the transformation
$$
g^{i0}\rightarrow g'^{i0}=\frac{\partial x'^i}{\partial x^j}g^{j0}+\frac{\partial x'^i}{\partial t}g^{00}\;
$$
We can evidently choose the time-dependence of  $x'^i({\bf x},t)$ so that $g'^{i0}=0$, by solving the differential equation 
$$ \frac{\partial x'^i}{\partial t}=-\frac{\partial x'^i}{\partial x^j}g^{j0}/g^{00}$$
for any arbitrary choice of  $x'^i({\bf x},t_0)$ at an initial time $t_0$.  In this case, also $g'_{i0}=0$.  Though not unique, this choice of space and time coordinates is clearly consistent with (1)---(3).  It still leaves us free to make purely spatial time-independent coordinate transformations, a freedom we have used in the arguments above.

These adiabatic solutions to the non-linear field equations far outside the horizon may look unfamiliar to readers who are familiar with the form of the adiabatic solution for scalar modes in the linear approximation in Newtonian gauge, for which $g_{i0}=0$ and $g_{ij}\propto \delta_{ij}$.  In [2] it is shown in the linear approximation that in the adiabatic mode  in Newtonian gauge there are perturbations to the matter fields $\chi_n(x)$ that do not vanish for large $a(t)$, and a perturbation to $\tilde{g}_{ij}$ that does not become time-independent in this limit:
$$ \delta\chi_n({\bf x},t)=-\frac{\zeta({\bf x})\dot{\bar{\chi}}_n(t)}{a(t)}\int_T^t a(t')\,dt'\;,~~~~~~\delta\tilde{g}_{ij}=2\delta_{ij}\zeta({\bf x})\left[1-\frac{H(t)}{a(t)}\int_T^t a(t')\,dt'\right]\;,$$
with $\zeta({\bf x})$ an infinitesimal  function only of position, and $T$ arbitrary.  But it is easy to see that by a re-definition of the space and time coordinates, we can  make all $\delta\chi_n$ vanish, keep $g_{i0}$ equal to zero, and make  $\delta\tilde{g}_{ij}$ equal to 
$$
\delta\tilde{g}_{ij}=2\delta_{ij}\zeta({\bf x})+2\frac{\partial^2\zeta({\bf x})}{\partial x^i\partial x^j}\int_T^t\frac{dt'}{a^3(t')}\int_T^{t'}a(t'')\,dt''\;,
$$
which in the limit of large $a(t)$ approaches the time-independent function $2\delta_{ij}\zeta({\bf x})$, with a correction of order $a^{-2}$.  In the non-linear case there is no advantage to using something like Newtonian gauge (for instance, by choosing space coordinates so that $\tilde{g}_{ij}=e^{2\zeta}[e^\gamma]_{ij}$ where $\partial_i\gamma_{ij}=0$ as well as $\gamma_{ii}=0$, as in [3]), and such a choice has the disadvantage of spoiling three-dimensional coordinate invariance.

This analysis allows us to make a rough estimate of the expected corrections to the constancy of the metric correlations and to the vanishing of the correlation functions for matter perturbations following single-field inflation.  At horizon exit the rate of change of the correlation functions of $\tilde{g}_{ij}$  is of order $H$, and from then until the end of inflation, the factor $1/a^2$ decreases by a factor roughly of order $e^{-120}$ to $e^{-140}$[14], so at the end of inflation we expect the  correlation functions of $\tilde{g}_{ij}$ to be changing at a rate of order $e^{-120}H$ to $e^{-140}H$.  Similarly, the correlation functions of $\tilde{g}_{ij}$ at the end of inflation are of the same order as at horizon exit, so with the decrease in $1/a^2$ we expect correlation functions for matter perturbations after inflation to less than the correlation functions of $\tilde{g}_{ij}$ by a factor of order $e^{-120}$ to $e^{-140}$.  This is the suppression factor that has to be overcome in order for physical processes during reheating like parametric amplification[12] to produce significant changes in observable correlation functions.

\begin{center}
{\bf III. Explicit Solutions for  Multiscalar Theories}
\end{center}

The arguments of the previous section were rather abstract, so to see  in more detail how they work out in practice, let us consider a  more concrete but still fairly general model.  To represent the matter fields, we suppose that there is a set of scalar  fields $\chi_n$, with a conventional kinematic Lagrangian and a completely arbitrary real potential $V(\chi)$.  The unperturbed values of these fields are functions $\bar{\chi}_n(t)$ of time that satisfy the field equations
\begin{equation}
\ddot{\bar{\chi}}_n+3H\dot{\bar{\chi}}_n+\frac{\partial V(\bar{\chi})}{\partial\bar{\chi}_n}=0
\end{equation}
where (in units with $8\pi G=1$) 
\begin{equation}
3H^2=\frac{1}{2}\sum_n\dot{\bar{\chi}}_n^2+V(\bar{\chi})\;.
\end{equation}
For the metric, we use the ADM parameterization[15]
\begin{eqnarray}
&& g_{00}=-N^2+g_{ij}N^iN^j\;,~~~~g_{0i}=g_{ij}N^j\equiv N_i\nonumber\\
&& g^{00}=-N^{-2}\;,~~~~g^{i0}=N^i/N^2\;,~~~~g^{ij}={}^{(3)}g^{ij}-N^iN^j/N^2\;,
\end{eqnarray}
where ${}^{(3)}g^{ij}$ is the reciprocal of the $3\times 3$-matrix $g_{ij}$.  
It will be convenient also to write 
\begin{equation}
g_{ij}({\bf x},t)=a^2(t)\tilde{g}_{ij}({\bf x},t)\;,
\end{equation}
where $a(t)$ is the Robertson--Walker scale factor, satisfying $\dot{a}/a=H$, with $H$ given by Eq,~(6),
The quantities $N$ and $N^i$ are auxiliary fields, whose time derivatives do not appear in the Lagrangian.  The Lagrangian for this theory is
\begin{eqnarray}
&& L=\frac{1}{2}\int d^3x\;\sqrt{-{\rm Det}g}\left\{-R^{(4)}-g^{\mu\nu}\sum_n\partial_\mu\chi_n\partial_\nu\chi_n-2V(\chi)\right\}\nonumber\\&&~~=\frac{a^3}{2}\int d^3x\;N\sqrt{\tilde{g}}\Bigg\{-a^{-2}\tilde{g}^{ij}\tilde{R}_{ij}+C^i{}_jC^j{}_i-(C^i{}_i)^2\nonumber\\&&~~~
+N^{-2}\sum_n\Big(\dot{\chi}_n-N^i\partial_i\chi_n\Big)^2-a^{-2}\tilde{g}^{ij}\sum_n\partial_i\chi_n\partial_j\chi_n-2V(\chi)\Bigg\}\;,~~
\end{eqnarray}
where $\tilde{g}^{ij}({\bf x},t)$ is the reciprocal of the matrix $\tilde{g}_{ij}({\bf x},t)$; $\tilde{g}({\bf x},t)$  is the determinant of $\tilde{g}_{ij}({\bf x},t)$; $\tilde{R}_{ij}({\bf x},t)$ is the three-dimensional Ricci tensor (with the sign convention of [2]) for the metric $\tilde{g}_{ij}({\bf x},t)$; and $C^i{}_j({\bf x},t)$ is the extrinsic curvature of the surfaces of fixed time
\begin{equation}
C^i{}_j\equiv a^{-2}\tilde{g}^{ik}\,C_{kj}\;,~~~~C_{ij}\equiv \frac{1}{2N}\Big[2a\dot{a}\tilde{g}_{ij}+a^2\dot{\tilde{g}}_{ij}-\tilde{\nabla}_i N_j-\tilde{\nabla}_j N_i\Big]\;,
\end{equation}
where $\tilde{\nabla}_i$ is the three-dimensional covariant derivative calculated with the three-metric $\tilde{g}_{ij}$.  
For future use, we also note the well-known relations (for $8\pi G=1$):
\begin{equation}
\sum_n\dot{\bar{\chi}}_n^2=-2\dot{H}\:~~~~~V(\bar{\chi})=3H^2+\dot{H}\;.
\end{equation}
Models of this sort  can be used both as  fairly realistic theories of inflation, and also as  surrogates for a theory of the matter and radiation after inflation.  Because we are allowing any number of scalar fields, this  model  will in general have solutions in which neither the perturbations to matter fields nor the rate of change of $\tilde{g}_{ij}$  vanish at late times, so it is not trivial to see that there is also an adiabatic solution in which they do go to zero at late time, and that this is the solution that is excited if during inflation there is only one non-negligible scalar field. 

The gravitational field equations derived from this Lagrangian are
\begin{equation}
\tilde{\nabla}_i\Big(C^i{}_j-\delta^i{}_jC^k{}_k\Big)=\frac{1}{N}\sum_n\partial_j\chi_n\Big(\dot{\chi}_n-N^i\partial_i\chi_n\Big)\;,
\end{equation}
\begin{eqnarray}
&&N^2\Big[-a^{-2}\tilde{g}^{ij}\tilde{R}_{ij}-C^i{}_jC^j{}_i+(C^i{}_i)^2-2V(\chi)\Big]\nonumber\\&&~~~=\sum_n(\dot{\chi}_n-N^i\partial_i\chi_n)^2+N^2a^{-2}\tilde{g}^{ij}\sum_n\partial_i\chi_n\partial_j\chi_n\;,
\end{eqnarray}
\begin{eqnarray}
&&\tilde{R}_{ij}-C^k{}_kC_{ij}+2C_{ik}C^k{}_j+N^{-1}\Big(-\dot{C}_{ij}+C^k{}_i\tilde{\nabla}_jN_k+C^k{}_j\tilde{\nabla}_iN_k\nonumber\\
&&~~~+N^k\tilde{\nabla}_kC_{ij}+\tilde{\nabla}_i\tilde{\nabla}_jN\Big)=-a^2\tilde{g}_{ij}V(\chi)-\sum_n\partial_i\chi_n\,\partial_j\chi_n\;,
\end{eqnarray}
and the scalar field equations are
\begin{eqnarray}
&&\frac{\partial}{\partial t}\left(\frac{\sqrt{\tilde{g}}}{N}\Big(\dot{\chi}_n-N^i\partial_i\chi_n\Big)\right)+\frac{3H\sqrt{\tilde{g}}}{N}\Big(\dot{\chi}_n-N^i\partial_i\chi_n\Big)
\nonumber\\&&~~=\frac{1}{a^2}\frac{\partial}{\partial x^i}\Big(\sqrt{\tilde{g}}\,N\tilde{g}^{ij}\partial_j\chi_n\Big)
+\frac{\partial}{\partial x^j}\left(\frac{\sqrt{\tilde{g}}\,N^j}{N}\Big(\dot{\chi}_n-N^i\partial_i\chi_n\Big)\right)
\nonumber\\&&~~~~~~~-\sqrt{\tilde{g}}N\frac{\partial V(\chi)}{\partial \chi_n}\;.
\end{eqnarray}
In line with the remarks of the  previous section, we look for a solution in which $\delta\chi_n({\bf x},t)\equiv \chi_n({\bf x},t)- \bar{\chi}_n(t)$ as well as $\dot{\tilde{g}}_{ij}$ and $\delta N\equiv N-1$ are all of order $a^{-2}(t)$ at late time.  
For convenience, in accordance with remarks at the end of the previous section, we also adopt a definition of space coordinates for which $N^i=0$.  We can then  write 
\begin{equation}
C^i{}_j=H\delta^i{}_j+\xi^i{}_j\;,
\end{equation}
where $\xi^i{}_j$ is, like $\dot{\tilde{g}}_{ij}$ and $\delta N$, a quantity whose leading term is of order $a^{-2}$:
\begin{equation}
\xi^i{}_j=\frac{1}{2}\tilde{g}^{ik}\left[\dot{\tilde{g}}_{kj}-2H\,\delta N\,\tilde{g}_{kj}\right]+O(1/a^4)\;.
\end{equation}
 Then the gravitational field equations (12)--(14) become
\begin{equation}
\tilde{\nabla}_i\Big(\xi^i{}_j-\delta^i{}_j\xi^k{}_k\Big)=\partial_j\sum_n\delta\chi_n\dot{\bar{\chi}}_n+O(a^{-4})\;,
\end{equation}
\begin{equation}
4\dot{H}\delta N=-a^{-2}\tilde{g}^{ij}\tilde{R}_{ij}+4H\xi^k{}_k-2\sum_n\dot{\bar{\chi}}_n\,\delta\dot{\chi}_n-2
\sum_n\frac{\partial V(\bar{\chi})}{\partial\bar{\chi}_n}\delta\chi_n+O(a^{-4})\;,
\end{equation}
\begin{equation}
\dot{\xi}^i{}_j+3H\xi^i{}_j+H\delta^i{}_j\xi^k{}_k-\dot{H}\delta N\delta^i{}_j=a^{-2}\tilde{g}^{ik}\tilde{R}_{kj}+\delta^i{}_j\sum_n\frac{\partial V(\bar{\chi})}{\partial \bar{\chi}_n}\delta\chi_n+O(a^{-4})\;.
\end{equation}
Using Eqs.~(19) and (5), we can rewrite Eq.~(20) in the form:
\begin{equation}
\dot{\Xi}^i{}_j+3H\Xi^i{}_j=\frac{1}{a^2}\left[\tilde{g}^{ik}\tilde{R}_{kj}-\frac{1}{4}\delta^i{}_j\tilde{g}^{kl}\tilde{R}_{kl}\right] +O(a^{-4})\;,
\end{equation}
where
\begin{equation}
\Xi^i{}_j\equiv\xi^i{}_j+\frac{1}{2}\delta^i{}_j\sum_n\dot{\bar{\chi}}_n\delta\chi_n\;.
\end{equation}
Also, Eq.~(18) now reads simply
\begin{equation}
\tilde{\nabla}_i\Big(\Xi^i{}_j-\delta^i{}_j\Xi^k{}_k\Big)=O(a^{-4})\;,
\end{equation}
Eq.~(21) has a solution:
\begin{equation}
\Xi^i{}_j({\bf x},t)=\left[{\cal G}^{ik}({\bf x}){\cal R}_{kj}({\bf x})-\frac{1}{4}\delta^i{}_j{\cal G}^{kl}({\bf x}){\cal R}_{kl}({\bf x})\right]\frac{1}{a^3(t)}\int_{T}^t a(t')\,dt'+\frac{B^i{}_j({\bf x})}{a^3(t)}+O(a^{-4})\;,
\end{equation}
where $T$ is any fixed time, ${\cal G}_{ij}({\bf x})$ is the value of $\tilde{g}_{ij}({\bf x},t)$ at that time, ${\cal R}_{ij}({\bf x})$ is the Ricci tensor calculated from the 3-metric ${\cal G}_{ij}({\bf x})$, and $B^i{}_j({\bf x})$ is a time-independent function of co-moving position.  It is convenient to choose $T$ at around the end of inflation, where $k/aH$ is smallest, in which case all terms in Eq.(24) are very small from soon after horizon exit to just before horizon re-entry.    The leading term in (24)  automatically satisfies Eq.~(23) because of the Bianchi identity satisfied by ${\cal R}_{ij}$.

To solve for the metric, we need to complete our choice of gauge.  By using Eqs.~(22), (17), (19), and (5), we have
\begin{equation}
\dot{\tilde{g}}_{ij}=2\tilde{g}_{ik}\Xi^k{}_j+\frac{2H^2}{\dot{H}}\tilde{g}_{ij}\Xi^k{}_k-\frac{H}{2a^2\dot{H}}\tilde{g}_{ij}\tilde{g}^{kl}\tilde{R}_{kl}-\tilde{g}_{ij}X+O(a^{-4})\;,
\end{equation}
where $X=O(1/a^2)$ arises from the perturbation to the scalar fields
\begin{equation}
X\equiv \sum_n\dot{\bar{\chi}}_n\delta\chi_n+\frac{H}{\dot{H}}\sum_n\Big(\dot{\bar{\chi}}_n\delta\dot{\chi}_n-\ddot{\bar{\chi}}_n\delta\chi_n\Big)\;.
\end{equation}
Under a shift $t\rightarrow t+\epsilon({\bf x},t)$ in the time coordinate, with $\epsilon$ of order $1/a^2$ (and a corresponding transformation $x^i\rightarrow x^i+{\cal G}^{ij}\int dt\,a^{-2}\partial \epsilon/\partial x^j$ to keep $N_i=0$), the perturbations to the scalar fields undergo the gauge transformation $\delta\chi_n({\bf x},t)\rightarrow \delta\chi_n({\bf x},t)-\epsilon({\bf x},t)\dot{\bar{\chi}}_n(t)$ to order $1/a^2$.  Hence, using Eq.~(11), to this order
\begin{equation}
X\rightarrow X+2\frac{\partial}{\partial t}\Big(\epsilon H\Big)\;,
\end{equation}
so we can evidently choose $\epsilon$ to make $X=0$.  This choice provides the great advantage that we can solve Eq.~(25) for $\tilde{g}_{ij}$ without first solving the field equations for the matter fields:
\begin{eqnarray}
&&\tilde{g}_{ij}({\bf x},t)={\cal G}_{ij}({\bf x})+2\left[{\cal R}_{ij}({\bf x})-\frac{1}{4}{\cal G}_{ij}({\bf x}){\cal G}^{kl}({\bf x}){\cal R}_{kl}({\bf x})\right]\int^t_T\frac{dt'}{a^3(t')}\int_{T}^{t'} a(t'')\,dt''\nonumber\\&&
~~~+\frac{1}{2}{\cal G}_{ij}({\bf x}){\cal G}^{kl}({\bf x}){\cal R}_{kl}({\bf x})\int_T^t\frac{H^2(t')\,dt'}{\dot{H}(t')a^3(t')}\int_{T}^{t'} a(t'')\,dt''\nonumber\\
&&~~~-\frac{1}{2}{\cal G}_{ij}({\bf x}){\cal G}^{kl}({\bf x}){\cal R}_{kl}({\bf x})\int_T^t \frac{H(t')\,dt'}{a^2(t')\,\dot{H}(t')}\nonumber\\&&
~~~+2\,{\cal G}_{ik}({\bf x})B^k{}_j({\bf x})\int_T^t \frac{dt'}{a^3(t')}+2\,{\cal G}_{ij}({\bf x})B^k{}_k({\bf x})\int_T^t \frac{H^2(t')\,dt'}{a^3(t')\,\dot{H}(t')}\nonumber\\&&~~~+O\Big(a^{-4}(t)\Big)\;,
\end{eqnarray}
where  $T$ is again some fixed time, conveniently chosen as the time at the end of inflation, and   ${\cal G}_{ij}({\bf x})$ and ${\cal R}_{ij}({\bf x})$ are the values of $\tilde{g}_{ij}$ and the associated Ricci tensor at that time.
This confirms that while far outside the horizon, the time-dependent part of $\tilde{g}_{ij}$ is of order $a^{-2}$.  But in the radiation or matter-dominated era the second, third, and fourth  terms in Eq.~(28) increase like $1/a^2H^2$, which produces the breakdown in these approximations when  physical wavelengths re-enter the horizon.

It remains to consider the scalar fields.  By using Eq.~(5) again, we can put the field equation (15) in the form
\begin{equation}
\delta\ddot{\chi}_n+3H\delta\dot{\chi}_n+\sum_m\frac{\partial^2V(\bar{\chi})}{\partial\bar{\chi}_n\partial\bar{\chi}_m}
\delta\chi_m=
\left(-\frac{1}{2}\tilde{g}^{ij}\dot{\tilde{g}}_{ij}+\delta\dot{N}+6H\delta N\right)\dot{\bar{\chi}}_n+2\delta N\ddot{\bar{\chi}}_n+O(1/a^4)\;.
\end{equation}
With $X=0$, we now have
\begin{equation}
\delta N=-\frac{1}{4\dot{H}a^2}\tilde{g}^{ij}\tilde{R}_{ij}+\frac{H}{\dot{H}}\Xi^k{}_k+\frac{1}{2H}\sum_n\dot{\bar{\chi}}_n\,\delta\chi_n+O(1/a^4)\;.
\end{equation}
Using Eqs.~(21), (25), and (30), we can put Eq.~(29) in the form
\begin{eqnarray}
&&\delta\ddot{\chi}_n+3H\delta\dot{\chi}_n-\frac{1}{2H}\sum_m \dot{\bar{\chi}}_n\dot{\bar{\chi}}_m\delta\dot{\chi}_m\nonumber\\&&~~~~~~+\sum_m\Bigg[\frac{\partial^2V(\bar{\chi})}{\partial\bar{\chi}_n
\partial\bar{\chi}_m}-\left(3-\frac{\dot{H}}{2H^2}\right)\dot{\bar{\chi}}_n\dot{\bar{\chi}}_m-\frac{1}{2H}(\dot{\bar{\chi}}_n\ddot{\bar{\chi}}_m
+2\ddot{\bar{\chi}}_n\dot{\bar{\chi}}_m)\Bigg]\delta\chi_m
\nonumber\\&&
~~~=\frac{1}{\dot{H}^2}\Big(\ddot{H}\dot{\bar{\chi}}_n-2\dot{H}\ddot{\bar{\chi}}_n\Big)\left(\frac{1}{4a^2}{\cal G}^{ij}{\cal R}_{ij}-H\Xi^k{}_k\right)+O(a^{-4})\;,
\end{eqnarray}
or, using Eq.~(24) for $\Xi$:
\begin{eqnarray}
&&\delta\ddot{\chi}_n+3H\delta\dot{\chi}_n-\frac{1}{2H}\sum_m \dot{\bar{\chi}}_n\dot{\bar{\chi}}_m\delta\dot{\chi}_m\nonumber\\&&~~~~~~+\sum_m\Bigg[\frac{\partial^2V(\bar{\chi})}{\partial\bar{\chi}_n
\partial\bar{\chi}_m}-\left(3-\frac{\dot{H}}{2H^2}\right)\dot{\bar{\chi}}_n\dot{\bar{\chi}}_m-\frac{1}{2H}(\dot{\bar{\chi}}_n\ddot{\bar{\chi}}_m
+2\ddot{\bar{\chi}}_n\dot{\bar{\chi}}_m)\Bigg]\delta\chi_m
\nonumber\\&&
~~~=\frac{1}{\dot{H}^2}\Big(\ddot{H}\dot{\bar{\chi}}_n-2\dot{H}\ddot{\bar{\chi}}_n\Big)\left[{\cal R}({\bf x})\left(\frac{1}{4a^2}-\frac{H}{4a^3}\int_T^t a(t')\,dt'\right)-\frac{H\,B^i{}_i({\bf x})}{a^3}\right]\nonumber\\&&~~~~~~+O(a^{-4})\;.
\end{eqnarray}

An inhomogeneous differential equation of this form will have a solution in which a non-zero curvature scalar ${\cal R}$ will generate perturbations of order $1/a^2$ in the various scalar fields, as anticipated in the previous section.  The field equations also have isocurvature solutions  in which ${\cal R}=0$ and  there are small perturbations to the scalar field, not necessarily of order $1/a^2$, for which
\begin{eqnarray}
&&0=\delta\ddot{\chi}_n+3H\delta\dot{\chi}_n-\frac{1}{2H}\sum_m \dot{\bar{\chi}}_n\dot{\bar{\chi}}_m\delta\dot{\chi}_m\nonumber\\&&+\sum_m\Bigg[\frac{\partial^2V(\bar{\chi})}{\partial\bar{\chi}_n
\partial\bar{\chi}_m}-\left(3-\frac{\dot{H}}{2H^2}\right)\dot{\bar{\chi}}_n\dot{\bar{\chi}}_m-\frac{1}{2H}(\dot{\bar{\chi}}_n\ddot{\bar{\chi}}_m
+2\ddot{\bar{\chi}}_n\dot{\bar{\chi}}_m)\Bigg]\delta\chi_m\;,\nonumber\\&&{}
\end{eqnarray}
and $X=0$.   To tell what solutions actually describe the metric and matter of the universe, we need a specific model of inflation, such as single field inflation, to which we now turn.

\begin{center}
{\bf IV. Single-field Inflation, and its Aftermath}
\end{center}

During single-field inflation there is by assumption only one non-zero $\chi_n$, say $\chi_1$, so Eq.~(11) gives 
$\dot{H}=-\dot{\bar{\chi}}_1^2/2$ and $\ddot{H}=-\dot{\bar{\chi}}_1\ddot{\bar{\chi}}_1$, and we see that in this case the right-hand side of Eq.~(32) vanishes.  Thus during single-field inflation Eq.~(32) is a homogeneous differential equation for $\delta\chi_1$, and therefore allows a solution $\delta\chi_1=0$, which of course it must, since we can arrange that $\delta\chi_1=0$ by a choice of gauge consistent with the gauge choice $X=0$ used to derive Eq.~(32).    

This shows that the non-linear field equations for single-field inflation have a solution in which $\delta \chi_1=0$, and in which for late times  $\tilde{g}_{ij}({\bf x},t)$  is attracted to a time-independent metric ${\cal G}_{ij}({\bf x})$, with corrections of order $a^{-2}$ and $a^{-3}$ given by Eq.~(28).    We know by explicit calculation that in the linear approximation all solutions are in the basin of attraction for this asymptotic solution[16], but it is  difficult to show that the relevant solution of the full non-linear equations  is in this basin of attraction, and we shall simply assume that this is the case.

Then at the end of inflation the transfer of energy from the inflation turns on other scalar fields, and the right-hand side of Eq.~(32) becomes non-zero.  As we have seen the general solution for the scalar field perturbations is a forced term of order $a^{-2}$, plus a solution of the homogenous equation (33).  In general the solution of the homogeneous equation could be of any order in $a$, but by definition during single field inflation in our gauge all $\delta\chi_n$ and $\delta\dot{\chi}_n$ are negligible, and with these initial conditions the solution of the homogeneous equation must  be of order $a^{-2}$ to cancel the $O(a^{-2})$ terms in the solution of the inhomogeneous equation immediately after single field inflation.  Thus as expected, for this solution all perturbations to the matter fields become of order $1/a^2$ outside the horizon, and we have a pure adiabatic solution, with negligible corrections.

\begin{center}
{\bf V. Tree-Approximation Correlation Functions}
\end{center}

If the results we have obtained so far really applied to the metric and matter perturbations in the Heisenberg picture, we could conclude that with a suitable definition of coordinates, all correlation functions involving only $\tilde{g}_{ij}$ (or functions of $\tilde{g}_{ij}$) become time-independent outside the horizon, and that all correlation functions involving perturbations to $g_{00}$, $g_{0i}$, and matter variables become negligible outside the horizon.  But as mentioned in the Introduction, the presence of quantum fluctuations of arbitrarily small wave lengths invalidates the expansions in powers of $1/a$ as applied to the Heisenberg picture interacting fields.  To avoid this problem we must  limit our consideration to tree graphs for correlation functions, on  the assumption that the contributions of graphs with loops  are much smaller.  We can as usual apply the results of Sections II -- IV to the Heisenberg picture quantum fields, but calculate correlation functions only to lowest order in interactions to avoid loop graphs, hoping that this is a good approximation.  Here we want to consider an alternative approach, in which one explicitly considers only tree graphs.

In the appendix we review the general tree theorem of [11], which shows how to calculate the sum of tree graphs for  correlation functions by a solution of the {\em classical} field equations, subject to certain constraints.  To illustrate the use of this theorem, in this section we will apply this theorem  to the correlation functions of the reduced metric  $\tilde{g}_{ij}\equiv g_{ij}/a^2$  during single-field inflation, adopting space and time coordinates for which there is no perturbation to the inflaton field, and for which $g_{i0}=0$.

The generating function $W[J,t_1]$ for correlation functions of $\tilde{g}_{ij}$ at a time $t_1$ is defined by
Eq.~(A.1), which for this case takes the form
\begin{equation}
\exp\Big\{W[J,t_1]\Big\}\equiv \left\langle 0, {\rm in}\left|\exp\Big[\int d^3x\; \tilde{g}^H_{ij}({\bf x},t_1)J^{ij}({\bf x})\Big]\right|0,{\rm in}\right\rangle\;,
\end{equation}
where $\tilde{g}^H_{ij}({\bf x},t)$ is the Heisenberg-picture quantum mechanical operator corresponding to $\tilde{g}_{ij}({\bf x},t)$.  
Correlation functions for $\tilde{g}_{ij}({\bf x},t_1)$ are calculated according to Eq.~(A.2), which here reads
\begin{equation}
\langle 0, {\rm in}| \tilde{g}^H_{ij}({\bf x},t_1)\,\tilde{g}^H_{kl}({\bf y},t_1)\cdots|0,{\rm in}\rangle =\left[\frac{\partial^n }{\partial J^{ij}({\bf x})\,\partial J^{kl}({\bf y})\cdots}\exp\Big\{W[J,t_1]\Big\}\right]_{J=0}
\end{equation}
We want to evaluate $W[J,t_1]$ for late times during inflation, at which the Robertson--Walker scale factor $a(t_1)$ becomes exponentially large, from which we can calculate the late-time expectation value of products of the operators $\tilde{g}_{ij}$ at various space coordinates or wave numbers.

As described in the appendix, to calculate $W$ in the tree approximation we construct   complex  {\em c-number} metric fields $\tilde{g}_{ij}({\bf x},t)$  together with a complex auxiliary field $N({\bf x},t)$, satisfying the constraints:

\vspace{6pt}

\noindent
(A) The fields  
 satisfy the Euler--Lagrange equations.  In  the our case, they are  Eqs.~(12)--(14) with no scalar field perturbations.  

\vspace{6pt}

\noindent
(B) The fields $\tilde{g}^{ij}$  satisfy constraints at time $t_1$:
\begin{eqnarray}
&&{\rm Im}\,\tilde{g}_{ij}({\bf x},t_1)=0\;,\\
&&{\rm Im}\,\left\{\frac{\delta L[\tilde{g},\dot{\tilde{g}},t_1]}{\delta \dot{\tilde{g}}_{ij}({\bf x},t_1)}\right\}=-J^{ij}({\bf x})
\;.
\end{eqnarray}

\vspace{6pt}

\noindent
(C) $\tilde{g}_{ij}$  satisfies a positive frequency constraint at time $t\rightarrow -\infty$, that it behaves as  a superposition of terms proportional to $\exp(-i\omega t)$ , with $\omega$ various {\em positive} frequencies.

\vspace{6pt}

\noindent
These constraints give the functions $\tilde{g}_{ij}({\bf x},t)$  an implicit dependence on both the current  $J$ and on the time $t_1$ at which correlations are to be measured.  With the functions $\tilde{g}_{ij}({\bf x},t)$ and $N({\bf x},t)$ constructed in this way, the generating function is given by Eq.~(A.6), which here reads
\begin{equation}
W[J,t_1]_{\rm tree}=\int_{-\infty}^{t_1}{\rm Im}\,L[\tilde{g}(t),\dot{\tilde{g}}(t),t]\,dt+\int d^3x\; J^{ij}({\bf x}) \tilde{g}_{ij}({\bf x},t_1)\;.
\end{equation}

We showed in Section III that the non-linear field equations have  a solution for $\tilde{g}_{ij}({\bf x},t)$ that for late times is attracted to a time-independent metric ${\cal G}_{ij}({\bf x})$.    But as remarked  in the appendix, this is not enough to conclude that the correlation functions for $\tilde{g}_{ij}({\bf x},t)$ become time-independent at late time.  We must also show that the constraints (36) and (37) do not give  $\tilde{g}_{ij}({\bf x},t)$  any dependence on the time $t_1$ at which the constraints  are imposed, provided $a(t_1)$ is sufficiently large, and we must consider the convergence of the time integral in Eq.~(38) as $a(t_1)$ at the upper limit $t_1$ becomes large.   

For large $a(t_1)$, the constraint (36) simply provides the $t_1$-independent condition that the leading term 
${\cal G}_{ij}({\bf x})$ in Eq.~(28)  is real for all ${\bf x}$.  It follows then that the Ricci tensor ${\cal R}_{ij}({\bf x})$ for the metric ${\cal G}_{ij}({\bf x})$ is real, so the terms in (28)  of order $1/a^2$ are also real.  The leading terms in ${\rm Im}\,\tilde{g}_{ij}({\bf x},t_1)$ are then of order $1/a^3$:
\begin{eqnarray}
&& {\rm Im}\,\tilde{g}_{ij}({\bf x},t)=2{\cal G}_{ik}({\bf x})\,{\rm Im}B^k{}_j({\bf x})\,\int_T^t \frac{dt'}{a^3(t')}+2{\cal G}_{ij}({\bf x})\,{\rm Im}B^k{}_k({\bf x})\,\int_T^t\frac{H^2(t')\,dt'}{a^3(t')\,\dot{H}(t')}\nonumber\\&&~~~~~~~~~+O\Big(a^{-4}(t) \Big)\;.
\end{eqnarray}
The functional derivative appearing in the constraint (37) is
\begin{equation}
\frac{\delta L[\tilde{g}(t),\dot{\tilde{g}}(t),t]}{\delta \dot{\tilde{g}}_{ij}({\bf x},t)}=\frac{a^3(t)\,\sqrt{\tilde{g}({\bf x},t)}}{2}\tilde{g}^{ik}({\bf x},t)\Big(-2H(t)
\delta^j{}_k+\xi^j{}_k({\bf x},t)-\delta^j{}_k\,\xi^l{}_l({\bf x},t)\Big)\;.
\end{equation}
The metric is constrained by Eq.~(36) to be real at $t=t_1$, so the term $-2H\delta^i{}_j$ in parentheses makes a contribution to this functional derivative that is also real at $t=t_1$, but the tensor $\xi^i{}_j$ in the other two terms has an imaginary part given by the $O(a^{-3})$ term in Eq.~(24) (with $\Xi^i{}_j$ replaced with $\xi^i{}_j$, which in the absence of scalar field perturbations is the same):
\begin{equation}
{\rm Im}\,\xi^i{}_j({\bf x},t)=a^{-3}(t) {\rm Im}\,B^i{}_j({\bf x})+O\Big(a^{-4}(t)\Big)\;,
\end{equation}
 so                                                                                                                                                                   
\begin{equation}
{\rm Im}\frac{\delta L[\tilde{g}(t),\dot{\tilde{g}}(t)]}{\delta \dot{\tilde{g}}_{ij}({\bf x},t)}=\frac{\sqrt{{\cal G}({\bf x})}}{2}{\cal G}^{ik}({\bf x}){\rm Im}\Big(B^j{}_k({\bf x})-\delta^j{}_k\,B^l{}_l({\bf x})\Big)+O\Big(a^{-1}(t)\Big)\;.
\end{equation}
Thus the constraint (37) does become independent of  $t_1 $ for large $t_1$.  

This is not just a happy accident.  We can understand the asymptotic constancy of the left-hand side of Eq.~(37) by recalling the Euler--Lagrange equations
$$
  \frac{\partial}{\partial t}\frac{\delta L[\tilde{g}(t),\dot{\tilde{g}}(t)]}{\delta \dot{\tilde{g}}_{ij}({\bf x},t)}=\frac{\delta L[\tilde{g}(t),\dot{\tilde{g}}(t)]}{\delta \tilde{g}_{ij}({\bf x},t)}\;.$$
The imaginary part of the right-hand side decreases as $1/a^2$, so the  left-hand side of Eq.~(37) becomes constant for large $a$.

This leaves the question of the convergence of the integral over time in Eq.~(38) for large $a(t_1)$.  Let's first consider  the terms in the gravitational part of the Lagrangian (9) that contain either 0 or 1 space or time derivative.  Since $N$ is fixed by the condition that the Lagrangian be stationary in $N$, to first order in $\delta N$ we can set $N=1$.  It is then straightforward to calculate that the terms in the gravitational part of $L$ of zeroth or first order in derivatives add up to
\begin{equation}
L_1[\tilde{g},\dot{\tilde{g}},t]=\frac{a^3}{2}\int d^3x \sqrt{\tilde{g}}\,\Big[-12H^2-4\dot{H}-2H\tilde{g}^{ij}\dot{\tilde{g}}_{ij}-8H
 \tilde{g}^{ij}\tilde{\nabla}_iN_j\Big]\;.
\end{equation}
The final term in square brackets integrates to zero (and in any case vanishes for the choice we have made of spatial coordinates), leaving us (as already noted in [3]) with a total time derivative
\begin{equation}
L_1[\tilde{g},\dot{\tilde{g}},t]=-2\frac{d}{dt}\left(a^3H\int d^3x \sqrt{\tilde{g}}\right)
\end{equation}
As remarked in the appendix, such a total time derivative in the Lagrangian has no effect on the correlation functions.  This leaves the terms in the gravitational part of $L$ that are of second order in $\xi$ or that involve the space curvature.  According to Eq.~(39) the imaginary part of $N\sqrt{\tilde{g}}\tilde{g}^{ij}\tilde{R}_{ij}$ is of order $a(t)^{-3}$  at late time, which cancels the over-all factor $a^3$ in the Lagrangian, so this term  makes a contribution  of order $a^{-2}$.  According to Eq.~(41), the imaginary part of any  second-order function of the $\xi$ is of order $a^{-2}\times a^{-3}$, so again such terms make contributions to 
${\rm Im}(L-L_1)$ that at late times are of order $a^{-2}$.  The time integral in Eq.~(38) therefore converges to a finite limit for large $a(t_1)$ exponentially fast, as $\int^{t_1} a(t)^{-2}dt$.  This concludes the proof that in single field inflation the generating function $W[J,t_1]$ converges to a $ t_1$-independent function for large $ t_1$, and therefore so do the correlation functions of $\tilde{g}_{ij}$.

This demonstration, that  the correlation functions for the metric   converge to  $ t_1$-independent functions for large $ t_1$,  does  not imply that these limits are {\em uniform} in the coordinates appearing as arguments of the metric components.  
In fact, if we set coordinates equal, the correlation functions do not converge to finite limits.  For instance,  if to avoid ultraviolet divergences we define $\zeta_r(t)$ as the average of the curvature perturbation $\zeta({\bf x})$ over a very small co-moving volume $r^3$ whose physical radius $a(t)r$ over the times of interest remains much less than the horizon size $1/H(t)$, then it can be shown that in slow roll inflation the tree-approximation vacuum expectation value of $\zeta^2_r(t)$ increases like $a(t)^{n_S-1}$ for $n_S>1$ (where $n_S$ is the usual scalar mode slope parameter)  and like $\ln a(t)$ for $n_S=1$, though it does approach a constant  for $n_S<1$.  
Because of the way that $a(t)$ and the co-moving coordinates $x^i$ enter in the flat-space Robertson--Walker metric, they have no physical significance in themselves; it is only $a(t)$ times differences of  co-moving coordinates that have a significance, as physical separations.  Thus we expect the metric correlation functions to approach constant limits only    
when all such physical separations become large compared with the horizon size $1/H$.  Of course, in practice we are chiefly interested in the Fourier transforms of the correlation functions, in which case the physical wave numbers are the co-moving wave numbers divided by $a(t)$, and we expect these Fourier transforms to approach finite limits only when all physical wave numbers become much less than $H(t)$.

\vspace{12pt}

I am grateful for helpful conversations with Raphael Flauger, Eiichiro Komatsu, David Lyth, Juan Maldacena, Misao Sasaki, and Richard Woodard.  This material is based upon work supported by the National Science Foundation under Grant No. PHY-0455649 and with support from The Robert A. Welch Foundation, Grant No. F-0014

\vspace{12pt}

\setcounter{equation}{0}
\renewcommand{\theequation}{A.\arabic{equation}}

\begin{center}
{\bf  Appendix: The Tree Theorem}
\end{center}

In this appendix we will review the general tree theorem of [11], in a somewhat simplified notation, and add a remark that is needed in Section V.  We consider a general Lagrangian system, with Hermitian Heisenberg-picture canonical operators $q^H_a(t)$, and Lagrangian (not Lagrangian density) $L[q^H(t),\dot{q}^H(t),t]$, possibly with an intrinsic time dependence.  In field theories the index $a$ incorporates a space coordinate ${\bf x}$ as well as discrete indices labeling the various field components; a sum over $a$ includes an integral over ${\bf x}$ as well as sums over discrete indices; and derivatives with respect to $q_a(t)$ are interpreted as functional derivatives.  We wish to calculate the generating function $W[J,t_1]$, a real function of a set of real c-number currents $J_a$, which is defined by
\begin{equation}
\exp\Big\{W[J,t_1]\Big\}\equiv \left\langle 0, {\rm in}\left|\exp\Big[\sum_a q^H_a(t_1)J_a\Big]\right|0,{\rm in}\right\rangle\;,
\end{equation}
where   $|0,{\rm in}\rangle$ is a state defined to look like the vacuum state at an early time, which in this paper we take as $t=-\infty$.  From $W$ we can calculate expectation values in this state of products of any number $n$ of $q^H$s at the time $t_1$:
\begin{equation}
\langle 0, {\rm in}| q^H_a(t_1)\,q^H_b(t_1)\cdots|0,{\rm in}\rangle =\left[\frac{\partial^n }{\partial J_a\,\partial J_b\cdots}\exp\Big\{W[J,t_1]\Big\}\right]_{J=0}
\end{equation}
To calculate $W[J,t_1]$ in the tree approximation, we construct  {\em complex} c-number  functions $q_a(t)$, subject to three conditions:

\vspace{6pt}

\noindent
(A) The $q_a(t)$ 
 satisfy the Euler--Lagrange equations
\begin{equation}
\frac{\partial}{\partial t}\frac{\partial L[q(t),\dot{q}(t),t]}{\partial\dot{q}_a(t)} =\frac{\partial L[q(t),\dot{q}(t),t]}{\partial q_a(t)}\;.
\end{equation}
(In extending the Lagrangian to complex variables, we take it as a real function, in the sense that 
$L^*[q(t),\dot{q}(t),t]=L[q^*(t),\dot{q}^*(t),t]$.)

\vspace{6pt}

\noindent
(B) The $q_a(t)$ satisfy  constraints at time $t_1$:
\begin{eqnarray}
&&{\rm Im}\,q_a(t_1)=0\;,\\
&&{\rm Im}\,\frac{\partial L[q(t_1),\dot{q}(t_1),t_1]}{\partial \dot{q}_a(t_1)}=-J_a\;.
\end{eqnarray}

\vspace{6pt}

\noindent
(C) The $q_a(t)$ also satisfy a positive frequency constraint at time $t\rightarrow -\infty$, that they behave as  superpositions of terms with time-dependence  $\exp(-i\omega t)$, with $\omega$ various {\em positive} frequencies.

\vspace{6pt}

 \noindent
(In [11] the functions $q_a(t)$ were denoted $q_{La}(t)$; we are here taking advantage of the fact that for real currents, the other functions $q_{Ra}(t)$ introduced in [11] are just $q^*_{La}(t)$.)  The constraint (B) gives the $q_a(t)$ an implicit dependence on $t_1$ as well as on the $J_a$.  
With $q_a(t)$ calculated subject to these three constraints, the contribution of connected tree graphs to the generating function is given  by
\begin{equation}
W[J,t_1]_{\rm tree}=\int_{-\infty}^{t_1}{\rm Im}\,L[q(t),\dot{q}(t),t]\,dt+\sum_a J_a q_a(t_1)\;.
\end{equation}
We are concerned in this paper with the limit of $W$ for large $t_1$ (or more precisely, for large  $a(t_1)$).

From the foregoing, we can see that, in order to conclude that the generating function becomes independent of $t_1$ when $t_1$ is sufficiently large, it is not enough to show that the quantities $q_a(t_1)$ approach finite limits for large $t_1$.  We must also show that the integral in Eq.~(A.6) converges in this limit.  (There is no problem with the convergence at very early times, where the integrand oscillates increasingly rapidly.)  Further, because the constraints (A.4) and (A.5) are applied at time $t_1$, we must show that the quantities ${\rm Im}\{\partial L[q(t_1),\dot{q}(t_1),t_1]/\partial \dot{q}_a(t_1)\}$  as well as ${\rm Im}\,q_a(t_1)$ approach finite $t_1$-independent limits for large $t_1$.

In order to evaluate  the late time behavior of the correlation function in Section V, we  need to supplement this general review with a remark about the effect of adding to the Lagrangian a derivative term:
\begin{equation}
\Delta L(t)=\frac{d}{dt}F[q(t),t]\;,
\end{equation}
with $F[q(t),t]$ an arbitrary function of $t$ and of the $q_a(t)$, which is real in the same sense as $L$; that is, $F^*[q(t),t]=F[q^*(t),t]$.
It is familiar that such derivative terms do not matter in calculating the S-matrix, because there the Lagrangian enters in integrals over all time, but  in calculating the generating function here we need to integrate the Lagrangian only up to time $t_1$, and the Lagrangian also enters in the constraint (A.5).  Nevertheless, we can easily see that in calculating the generating function, as in S-matrix calculations, the change (A.7) has no effect.  First, adding a derivative term (A.7) obviously has no effect on the Euler-Lagrange equations (A.3).  The only other place where the Lagrangian enters in constructing the functions $q_a(t)$ is in the constraint (A.5), but adding the derivative term (A.7) changes   the left-hand side of Eq.~(A.5) by
\begin{equation}
\Delta\, {\rm Im} \frac{\partial L[q(t_1),\dot{q}(t_1),t_1]}{\partial \dot{q}_a(t_1)}={\rm Im} \frac{\partial F[q(t_1),t_1]}{\partial q_a(t_1)}\;,
\end{equation}
and this vanishes because the constraint (A.4) requires that $q_a(t_1)$ be real.  Hence the change (A.7) has no effect on the functions $q_a(t)$.  The only effect on the generating function (A.6) is then to change it by an amount
\begin{equation}
\Delta W[J,t_1]_{\rm tree}=\int_{-\infty}^{t_1}{\rm Im}\,\Delta L[q(t),\dot{q}(t),t]\,dt={\rm Im}F[q(t_1),t_1]
\end{equation}
and this vanishes because again the constraint (A.4) requires that $q_a(t_1)$ be real.

\begin{center}
{\bf References}
\end{center}

\begin{enumerate}

%1
\item For a review, see N. Bartolo, E. Komatsu, S. Matarrese, and A. Riotto, Phys. Rep. {\bf 402}, 103 (2004).

%2
\item For a review with references to the original literature, see S. Weinberg, {\em Cosmology} (Oxford University Press, 2008), Sec. 5.4.

%3
\item J. M. Maldacena, J. High Energy Phys. {\bf 05}, 013 (2003).

%4
\item D. H. Lyth, K. A. Malik, and M. Sasaki, J. Cosm. Astropart. Phys. {\bf 05}, 004 (2005).

%5
\item For a review and references to the original literature, see S. Weinberg, {\em The Quantum Theory of Fields}, Sec. 19.5 (Cambridge University Press, 1996).

\item For diverse discussions of this assumption in the context of the linear approximation see M. Sasaki and T. Tanaka, Prog. Theor. Phys. {\bf 99}, 763 (1998);  D. Wands, K. A. Malik, D. H. Lyth, and A. R. Liddle, Phys. Rev. D {\bf 62}, 043527 (2000);  A. R. Liddle and D. H. Lyth, {\em Cosmological Inflation and Large-Scale Structure} (Cambridge University Press, 2000);  D. H. Lyth and D. Wands, Phys. Rev. D {\bf 68}, 103516 (2003).   It  has been extended beyond the linear approximation in [4].

\item 
After this work was complete, I learned that the $O(a^{-2})$ and $O(a^{-3})$ terms in the metric in  the special case of single-field inflation have also been found by
Y. Tanaka and M. Sasaki, Prog. Theor. Phys. {\bf 181}, 455 (2007).  Their solution is different from that presented in Section III, presumably because they use a different gauge: In their gauge, the time is not defined to give the scalar field its unperturbed value, so that they find $O(a^{-2})$ and $O(a^{-3})$ perturbations to the scalar field, and the space coordinates are not defined to make $N^i=0$.

\item  S. Weinberg, Phys. Rev. D {\bf 72}, 043514 (2005).

%6
\item G `t Hooft and M. J. G. Veltman, Ann. Poincare Phys. Theor. {\bf A20}, 69 (1974); J. F. Donoghue, Phys. Rev. D {\bf 50}, 3874 (1994).

\item M. van der Meulen and J. Smit, J. Cosm. Astropart. Phys. {\bf 11},  023 (2007).

%7
\item S. Weinberg, Phys. Rev. D {\bf 78}, 063534 (2008).

%8
\item L. Kofman, A. Linde, and A.A.  Starobinsky, Phys. Rev. D {\bf 56}, 3258 (1997).  Non-Gaussianity due to parametric amplification is studied by K. Enqvist, A. Jokinen, A. Mazumdar, T. Multamaki, and A. Vaihkonen, J. Cosm. Astropart. Phys. {\bf 03}, 010 (2005); A. Jokinen and A. Mazumdar, J. Cosm. Astropart. Phys. {\bf 04}, 003 (2006); A. Chambers and A. Rajantie, Phys. Rev. Lett. {\bf 100}, 041302 (2008).

%9
\item For example, D. Seery and J. E. Lidsey, J. Cosm. Astropart. Phys. {\bf 06}, 0506 (2005); X. Chen, M-x. Huang, S. Kachru, and G. Shiu, J. Cosm. Astropart. Phys. {\bf 01}, 002 (2007); X. Chen, R. Easther, and E. A. Lim, J. Cosm. Astropart. Phys. {\bf 06}, 023 (2007) and 0801.3295.

\item A. R. Liddle and S. M. Leach, Phys. Rev. D {\bf 68}, 103503 (2003).

%10
\item R. S. Arnowitt, S. Deser, and C. W. Misner, in {\em Gravitation: An Introduction to Current Research}, ed. L. Witten (Wiley, New York, 1962): 227, now also  available as gr-qc/0405109.

%11
\item S. Weinberg, ref. [8], Eq.~(24).

%12

\end{enumerate}

\end{document}